# Characterizing Community Formation in Response to Extreme Weather Events through Human Mobility Networks

Cheng-Chun Lee[1, 3], Junwei Ma[1]*, Kai Yin[1], Siri Namburi[2], Xin Xiao[2], Ali Mostafavi[1]

[1] Urban Resilience.AI Lab, Zachry Department of Civil and Environmental Engineering, Texas A&M University, College Station, Texas, United States.

[2] Department of Computer Science and Engineering, Texas A&M University, College Station, Texas, United States.

3. Texas Department of Transportation, Austin, Texas, United States.

* Corresponding author: Junwei Ma, jwma@tamu.edu.

## ABSTRACT

Community formation in socio-spatial human networks is one of the important mechanisms for mitigating hazard impacts of extreme weather events. Research is scarce regarding latent network characteristics shaping community formation in human mobility networks during natural disasters. Here, we examined human mobility networks in Harris County, Texas, in the context of the managed power outage forced by 2021 Winter Storm Uri to detect communities and to evaluate latent characteristics in those communities. We examined three characteristics in the communities formed within human mobility networks: hazard-exposure heterophily, socio-demographic homophily, and social-connectedness strength. The results show that population movements were shaped by socio-demographic homophily, heterophilic hazard exposure, and social connectedness strength. Our results also indicate that a community encompassing more high-impact areas would motivate population movements to areas with weaker social connectedness. Our findings reveal important characteristics shaping community formation in human mobility networks in hazard response. Specific to managed power outages, formed communities are spatially co-located, underscoring a best management practice to avoid prolonged power outages among areas within communities, thus improving hazard exposure heterophily. The findings have implications for power utility operators to account for the characteristics of socio-spatial human networks when determining the patterns of managed power outages.

## 1. Introduction

Community formation is a fundamental property of complex networks, where nodes tend to cluster into subgroups with stronger internal connections than with the rest of the network. These communities capture meaningful intermediate-scale organization within complex systems, helping to simplify large networks, making them more interpretable, and revealing important structural patterns that may correspond to shared functions, behaviors, or contexts. Community detection supports a wide range of applications, including behavior prediction, anomaly detection, and targeted intervention. As a result, it has been widely applied across disciplines, such as identifying user groups in online social network [1-3] as well as uncovering spatial clusters of movement and multi-layer traffic structures in urban transportation systems [4-6].

Human mobility networks offer a particularly powerful lens for studying how populations respond and adapt to changing environmental or infrastructural conditions. In human mobility networks, nodes typically represent geographic locations, such as neighborhoods, census tracts, or regions, while edges represent the flow of people between them. Human mobility networks not only capture routine patterns of movement, such as commuting and daily activities, but also reflect shifts in behavior during disruptions, including evacuations, relocations, or shelter-seeking in response to crises. As a result, they have become essential tools in disaster research, offering insights into risk perception [7, 8], evacuation behavior [9, 10], and population recovery dynamics [11-13]. Community formation in human mobility networks during such events reveals how individuals reorganize spatially into groups of connected areas with shared exposure, vulnerability, or access to resources. Despite the growing use of mobility data in hazard studies, limited attention has been paid to understanding how these dynamic networks reorganize into communities during extreme events, and what these emergent structures reveal about population response and inequality. Addressing this gap, our study focuses on detecting and analyzing community formation in human mobility networks during a major infrastructure disruption caused by the 2021 Winter Storm Uri in Harris County, Texas.

Extreme weather events increasingly threaten critical infrastructure systems, often leading to managed interventions such as planned power outages to prevent catastrophic failure. During such disruptions, human mobility patterns can shift drastically as individuals seek access to safe, functional, or resource-rich areas. For instance, during the Winter Storm Uri in February 2021, prolonged subfreezing temperatures in Texas led to an unprecedented surge in electricity demand for heating, overwhelming the grid. To avoid a total system collapse, utility providers implemented rolling blackouts across the state, leaving nearly 4.5 million homes and businesses without power at the peak of the crisis [14]. In such events, managed power outages are intended to protect infrastructure, yet their effectiveness also depends on how well populations can adapt and respond. Understanding how communities form and reorganize through mobility during these outages is critical for improving future outage planning and risk communication strategies.

To this end, this study focuses on Harris County, Texas, one of the hardest-hit areas during Winter Storm Uri, and aims to characterize the formation of communities in human mobility networks during the managed power outage. Specifically, we investigate three key dimensions (shown in Fig.1) within these emergent communities: (1) hazard-exposure heterophily, to assess the extent to which areas with different levels of exposure are grouped together; (2) socio-demographic homophily, to examine whether communities reflect shared social characteristics such as race or income; and (3) social-connectedness strength, to evaluate the cohesion of movement within communities. By applying temporal community detection techniques to high-resolution mobility data, we aim to provide new insights into how population behavior and network structures evolve during infrastructure-related disruptions.

The remainder of the paper is organized as follows. Sections 2 reviews the existing literature and concludes the research gaps and questions. Section 3 details the data sources and methodological approach used in this study. Section 4 examines the three key community characteristics (i.e., hazard-exposure heterophily, socio-demographic homophily, and social-connectedness strength) observed in the identified communities. Section 5 discusses the contributions and implications of these characteristics for integrating socio-spatial network structures into disaster response planning and managed power outage management.

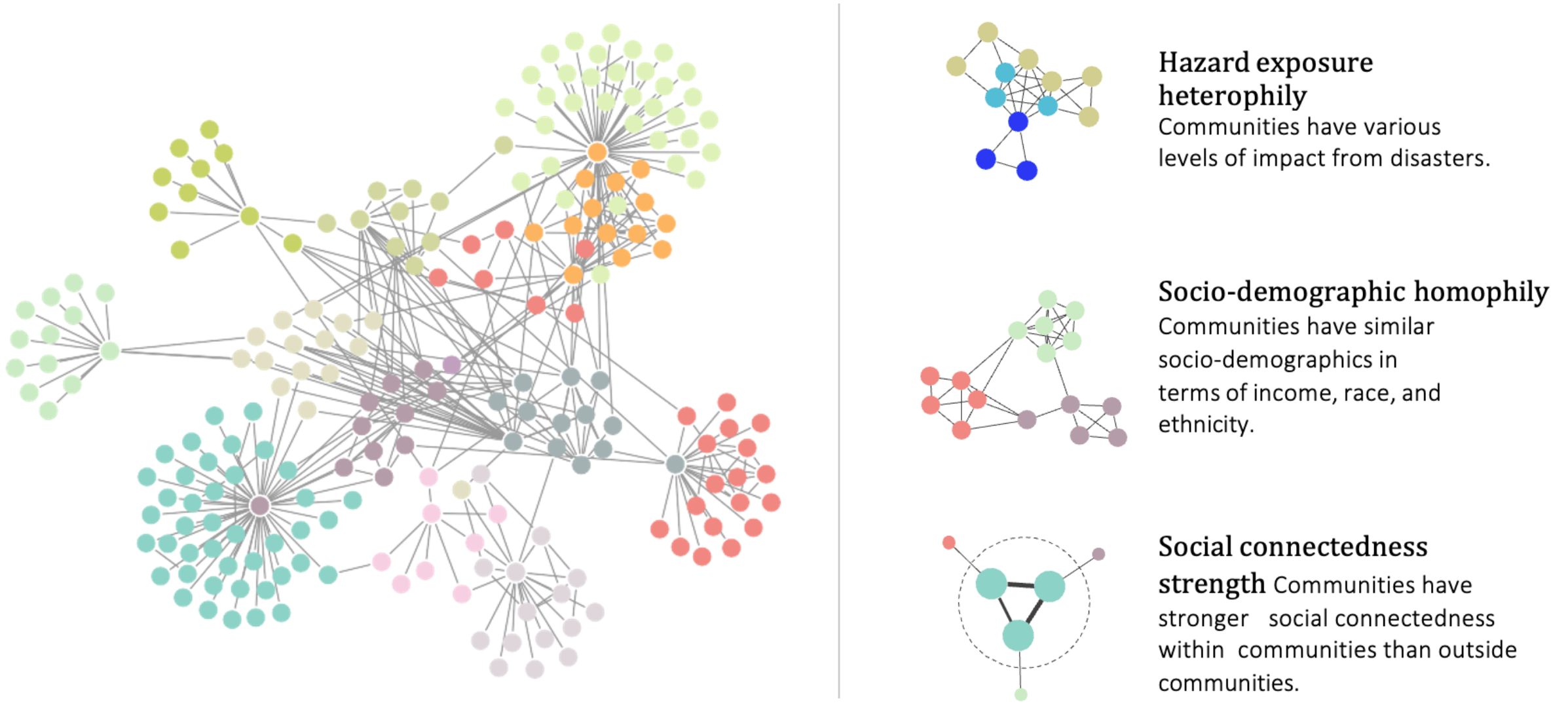

**Figure 1.** Schematic of community detection and the three important community characteristics.

## 2. Literature review

### 2.1 Big data observations of disaster-induced mobility

With the proliferation of mobile devices and location-based services, researchers now have access to high-resolution spatiotemporal data that captures the dynamics of population mobility in near real-time. The advent of location-aware devices and telemetry platforms has made it possible to observe human movement at unprecedented spatial and temporal granularity. This advancement has also proven valuable for studying mobility patterns in response to natural hazards, offering new insights into how populations behave before, during, and after disruptive events. For example, Li, Qiang [15] analyzes mobility data to detect evacuation patterns during Hurricane Ian, revealing spatial variation in compliance with evacuation orders and reinforcing the effectiveness of mandatory evacuation zones. Deng, Aldrich [9] reveals significant race and wealth disparities in evacuation patterns during Hurricane Harvey using large-scale GPS records. Li and Mostafavi [16] introduces a location intelligence framework to examine hurricane preparedness through visitation to point of interests including grocery store, gas station, pharmacies and home improvements before Hurricane Harvey. Ma and Mostafavi [17] tracks visitation to places and constructs human visitation networks to analyze fluctuations in latent lifestyle signatures during Hurricane Ida. Wang and Hu [18] implements a multiscale analysis of mobility network in southwest Florida during Hurricane Ian, revealing the nuances of mobility disruption and recovery across network scales. Hong, Bonczak [19] measures community resilience capacity using mobility shifts during Hurricane Harvey, revealing clear socioeconomic and racial disparities in both resilience and evacuation response. Yabe and Ukkusuri [20] assesses how income inequality influences evacuation, reentry, and spatial segregation in post-disaster scenarios. Despite these advancements, cascading cold weather events such as Winter Storm Uri remain significantly understudied in the mobility literature. Most existing research focuses on broad, statewide impacts while overlooking the fine-grained evolution of mobility patterns during the disruption and recovery phases. Moreover, a persistent gap in the literature is the tendency to analyze mobility data and power outage indicators in isolation, rather than in an integrated framework. This disconnection obscures how managed blackouts reshape daily origin-destination flows, constrain access to critical resources, and perturb the social and spatial structure of communities.

### 2.2 Dynamic community detection for networks

Community detection is a fundamental task in network science, aiming to uncover the hidden structure within networks by identifying clusters or groups of nodes that are more densely connected internally than with the rest of the network. The evolution of community detection algorithms has seen significant advancements over the years, moving from traditional methods to more sophisticated and dynamic approaches. Early community detection methods, such as the Girvan-Newman algorithm [21], relied on edge betweenness to iteratively remove edges and reveal community structure. Modularity-based approaches, introduced by Newman and Girvan [22], optimized the division of networks into communities by maximizing a modularity score. These methods, however, were computationally intensive and often struggled with large-scale networks. The advent of the Louvain method [23] marked a significant breakthrough, offering a more efficient approach to community detection by employing a multi-level optimization of modularity. This method quickly gained popularity due to its scalability and effectiveness in handling large networks.

Yet, real-world networks, such as human mobility networks, are frequently dynamic, combining both spatial and temporal dimensions. Static snapshots (e.g., a single day's mobility network) fail to capture the evolution of communities over time. Conventional multiplex approaches, where communities are detected independently in each snapshot and later aggregated, often overlook continuity and risk information loss. To more fully leverage temporal information, researchers have developed dynamic network representations including snapshot models[24, 25], link streams [26] and temporal networks [27]. Mucha, Richardson [28] introduced a multilayer framework linking the same node across time slices via interslice coupling, allowing stability and persistence of communities to be assessed over time. Greene, Doyle [29] proposed an approach that matches communities across consecutive snapshots to track evolution, and Appel, Cunha [27] used a shared factorization model to capture temporal, structural, and weighted graph properties. Qin, Li [30] later developed a density-based clustering algorithm tailored to detecting stable communities in dynamic graphs. Most notably, Boudebza, Cazabet [31] proposed a multi-scale method for detecting stable communities in link streams that combines change point detection with community detection to uncover community structures that remain robust across multiple temporal granularities. This method effectively identifies persistent communities without aggregating data into arbitrary time slices and has been validated using synthetic benchmarks and high-resolution, real-world temporal networks. Those advancements in dynamic community detection have provided an opportunity to more accurately capture the formation, evolution, and stability of communities in complex time-varying networks, enabling deeper insights into dynamic systems such as human mobility, information diffusion, and disaster response.

### 2.3 Homophily, heterophily, and social connectedness in hazard response

Building on the foundations of social mixing theories, this study investigates three key characteristics of mobility-based communities during a prolonged, managed power outage. Social mixing theory suggests that homophily, the tendency to associate with similar others, and heterophily, cross-exposure interactions with dissimilar others jointly shape collective behavior. In particular, Liu and Mostafavi [32] introduced the concept of hazard-exposure heterophily, referring to mobility links that connect spatial areas experiencing differing levels of hazard exposure. For example, if individuals in a highly impacted census tract maintain strong mobility ties with less-impacted areas, those ties can serve as vital pathways for temporary relocation or resource access. This heterophily is a latent property of socio-spatial networks that enhances communities' ability to respond to and recover from hazards. In addition, extensive research in both offline and digital contexts supports the idea that socio-demographic homophily, namely similarity in age, race, income, and other attributes, reinforces social ties and shapes behavior [33, 34]. During crises, such homophily can play a critical role in mobilizing support, as individuals are more likely to turn to those with whom they share identity and experience [35, 36]. A third and complementary dimension is social-connectedness strength, captured by metrics such as the social connectedness index, which reflects the density of digital friendships and has been used as a proxy for latent support networks that facilitate information flow and collective action. The social connectedness index could serve as a valuable proxy for latent support networks that enable the flow of information, aid, and collective action, particularly during crises. Research underscores that high levels of social cohesion and connectedness bolster community resilience, offering protective effects such as reducing psychological distress and facilitating recovery after disasters [37-39].

While each of these characteristics has been studied independently, no prior research has examined them simultaneously within dynamically detected mobility communities during an extreme hazard event. To address this gap, we implemented a temporal community detection algorithm to identify evolving local communities in human mobility networks and assessed the presence and interplay of these three characteristics during the response to a large-scale power outage.

## 3. Materials and methods

Figure 2 depicts the workflow of this study. We first collected data from four datasets (i.e., Spectus human mobility records, Mapbox population activity data, socio demographic attributes, and Meta social connectedness index) at the census tract level in Harris County, Texas. From these inputs, we constructed daily, directed origin-destination networks and computed activity density anomalies to classify power outage impact. We then applied a multi-scale, link-stream community detection algorithm to identify mobility communities that persist over time. Finally, we quantified three community-level metrics (i.e., hazard-exposure heterophily, socio-demographic homophily, and social-connectedness strength) to interpret how network structure and community characteristics shaped population response during Winter Storm Uri.

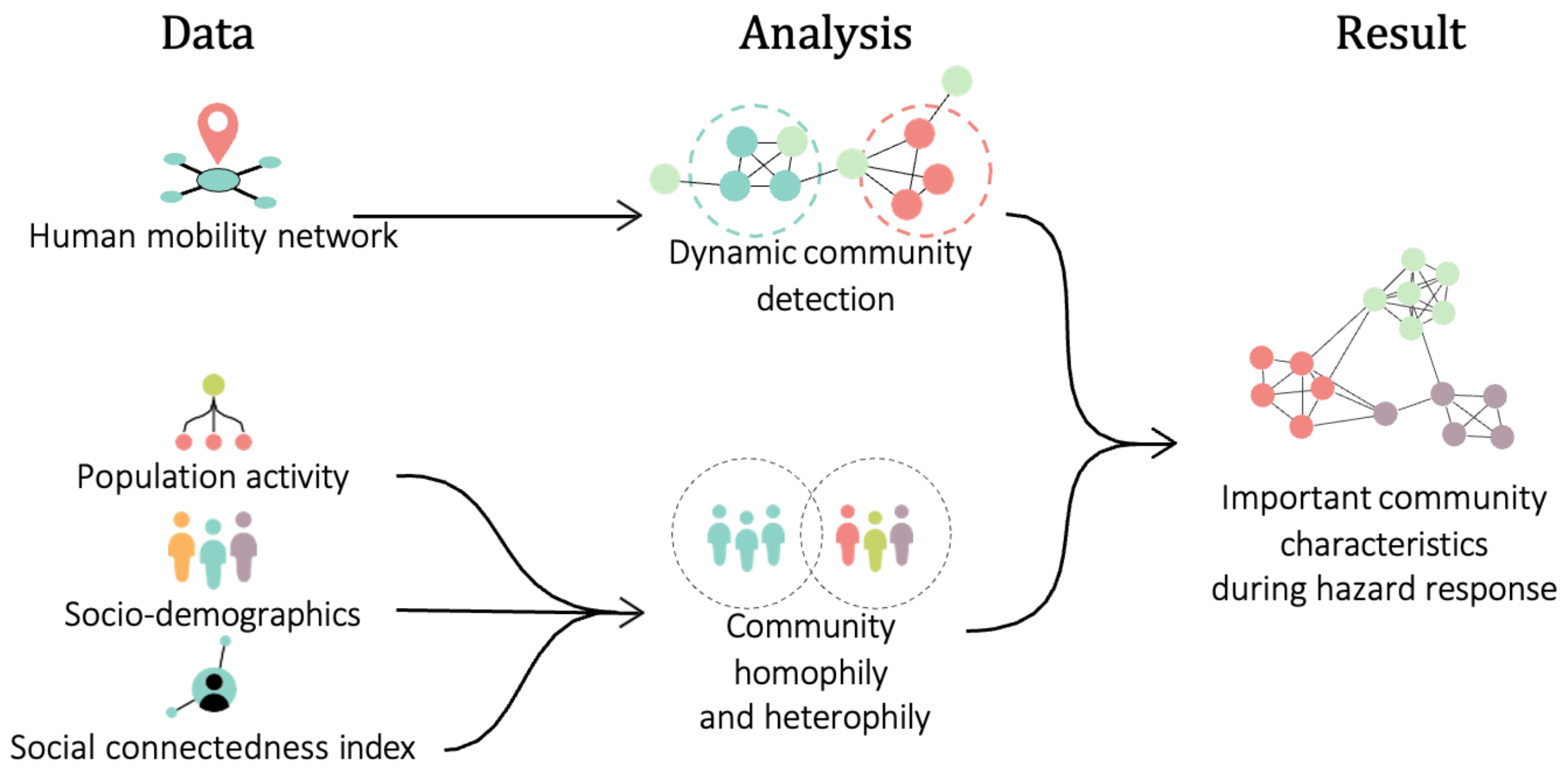

**Figure 2.** Overview of the research workflow.

### 3.1 Study area

This study collected and analyzed data from various sources in Harris County, Texas, which includes the Houston metropolitan area, during Winter Storm Uri, to examine important community characteristics under extreme weather responses. The Winter Storm Uri affected most of areas in North America from February 13 through February 17, 2021, and affected all of Texas, including Harris County, the study area, on the evening of February 14, 2021, bringing snow and below-freezing temperatures. Due to a lack of experience with cold weather, snow, or icy conditions, Texas residents were significantly impacted by the winter storm especially. The freezing weather broke historical records in both temperature and duration, and at least 146 people lost their lives in Texas due to hypothermia [14, 40, 41]. During the winter storm, the electricity infrastructure in Texas was overwhelmed due to heating demand, which led to the implementation of rolling power outages to avoid the collapse of the entire power grid in Texas. During the peak of this crisis, nearly 4.5 million homes and businesses in Texas lost power, and the estimated cost of the power outages was at least $195 billion USD. In particular, more than 90% of residents lost power in Harris County [41].

All datasets used in this study were aggregated to the census tract level and harmonized for comparability across sources. Harris County contains 786 census tracts, which serve as the unit of analysis for integrating mobility, population-activity, socio-demographic, and social-connectedness data.

## 3.2 Data sources

### 3.2.1 Human mobility data

The human mobility data used in this study is the location intelligence data obtained from Spectus. Traditionally, analyses corresponding to location data require prolonged data preparation and processing time using conventional surveys and adhering to strict privacy laws [42-44]. Spectus, a location data platform, provides access to privacy-enhanced GPS data of users who have opted to share their information, while thoroughly anonymizing their personal information through a CCPA- and GDPR-compliant framework. As of publication, Spectus has as many as 15 million active users in the United States [45]. This is made possible through the company's proprietary SDK present in its partner apps, which collect data anonymously. Spectus implements multiple safeguards beyond de-identification: home areas are obfuscated to the census block group level, sensitive points of interest are removed, and only aggregated outputs are permitted to leave the platform [45]. For this study, we accessed anonymized device-level fields (i.e., home census block group, stay census block group, and stay duration) and subsequently aggregated them to the census tract level (see Section 3.3.1 for network construction). No individual-level information was exported, and all analyses are reported at aggregate spatial units.

### 3.2.2 Population activity data

This study used telemetry-based population activity data provided by Mapbox to assess the impact of power outages. Due to the lack of publicly accessible power outage data, population activity data became a reliable proxy for understanding the extent of power outages [46, 47]. Mapbox aggregated and normalized raw active mobile phone counts by time period (e.g., day) and geographic unit (e.g., 100-meter by 100-meter spatial-resolution grid) to a baseline period of January 11 to 17, 2021. In particular, a scaling factor, activity index, was calculated by dividing the number of active mobile phones in each geographic unit by the 99.9th percentile for all geographic units for the baseline period. With the exclusion of possible outliers, the 99.9th percentile of mobile phone counts across all geographic units indicates the highest population activity in the study area. The activity index is further used to calculate the activity density in Section 3.3.2.

### 3.2.3 Socio-demographic data

The socio-demographic data used in this study was retrieved from the American Community Survey database, which is administrated by the U.S. Census Bureau. The American Community Survey regularly gathers information, including ancestry, educational attainment, income and earnings, disability status, and demographics such as age, race, and housing characteristics. This study used the 2019 five-year estimates data, which represents estimates over the five-year period from 2015 to 2019, at the census-tract level to understand the socio-demographics of population [48]. In particular, we obtained median household income, ratio of the Black population, and ratio of the Hispanic population for each census tract in Harris County. These measures provide socio-demographic context for the mobility communities identified in Section 3.3.3 and serve as inputs to the socio-demographic homophily metric described in Section 3.3.4.

### 3.2.4 Meta social connectedness index

The data of social connectedness index (SCI) is provided by Meta, a global networking service, as a part of its Data for Good program [49]. Facebook, one of Meta's platforms, has as many as 2.91 billion active users monthly from all over the world, of which around 180 million are from the United States. The SCI was developed by Bailey et al. [50], who used anonymized data of active Facebook user friendship ties between two geographical areas. By aggregating friendship links, the SCI is then calculated to measure the intensity of social connectedness between the locations in Section 3.3.4.

## 3.3 Methods

### 3.3.1 Human mobility network construction

To examine population responses during Winter Storm Uri, we harnessed the privacy-enhanced device-level mobility data provided by the Spectus platform, which includes the information of home census block groups, census block groups of stay, and stay durations for each anonymized device. In this study, we only retained stays having durations longer than two hours to account for the nature of relocation to cope with the impact. We then aggregated the data at the census tract level based on travel between home and stay census block groups along with counts of visits to construct origin-destination (OD) networks for each day in February 2021 in Harris County. Accordingly, a mobility network was constructed as a directed network. Consider a directed network represented as:

$$G = (V, E, w) \tag{1}$$

where $V$ is the set of nodes, $E$ is the set of edges that connects each of the nodes, and $w$ is the weight assigned to each of the edges. In this study, every node is the centroid of a census tract, and edges will be established if people are traveling from one census tract to another and $w$ is the trip counts between origins and destinations.

### 3.3.2 Characterization of the power outage impact

In this study, we calculated activity density at the census tract level per day to assess the impact level of power outages, using Mapbox's activity index (see Section 3.2.2). We compared the percentage change of activity density between the normal period (February 1 through February 7, 2021) and the impact period (February 8 through February 28, 2021), with larger percentage declines indicating greater outage impacts. For a given census tract *ct* and day *t*, activity density $Da(ct, t)$ is calculated from the activity index values $A_{u,t}$ of the *N* 100m × 100m grid cells $u$ within the tract:

$$Da(ct, t) = \sqrt{\frac{1}{N}\sum_{u=1}^{N} {A_{u,t}}^2} \quad (2)$$

where, $Da(ct, t)$ is the activity density in census tract *ct* at time *t*, $A_{u,t}$ is the normalized activity index for grid cell $u$ at time *t*, and *N* is the total number of grid cells within census tract *ct*.

The percent change in $Da$ between the normal and impact periods forms the basis for classifying tracts into high-, moderate-, and low-impact categories, which are subsequently used in the hazard-exposure heterophily metric (Section 3.3.4) and spatial analyses in Section 4.

### 3.3.3 Dynamic community detection

Community detection aims to identify subgroups of nodes within a network that are more densely connected to each other than to the rest of the network, thereby revealing latent structural relationships [1, 2]. In the context of human mobility networks, communities represent groups of locations between which movement is more frequent, offering insight into how people reorganize their travel patterns in response to extreme weather events and enabling examination of the factors shaping such spatial groupings. In this paper, we used the link stream-based algorithm proposed by Boudebza et al. [31] to detect stable communities in human mobility networks during Winter Storm Uri in February 2021 at multiple temporal scales. By applying such community detection technique to human mobility networks, we can identify communities where people have more connections in terms of their visit and stay activities. Based on the daily human mobility networks constructed $G_t = (V, E_t, \omega_t)$ in Section 3.3.1, we identified communities using a coarse-to-fine sliding-window strategy that captures both persistent and transient structures while reducing noise [31]. We formed a family of overlapping temporal windows (e.g., 7-, 5-, and 3-day windows) that advance by one day across February 2021. For each window $W$, we aggregated the daily OD flows by summing the edge weights $\omega_t(i, j)$ over all days $t \in W$, yielding a window-level directed, weighted network. On each such network, we applied the Louvain algorithm to obtain a static partition at that temporal scale [23], treating each window as a snapshot of the mobility system under a specific level of disruption.

To ensure that only well-formed communities are retained, we evaluated the cohesion of every detected community $C$ in window $W$ using conductance and its complement, inverse conductance. Let $\tilde{A}_{ij}^{(W)}$ denote the aggregated weight from tract $i$ to tract $j$ in window $W$, and let $V$ be the set of all tracts. We computed $\emptyset_w(C)$ and $I_w(C)$ as shown in Equation (3) and (4):

$$\emptyset_w(C) = \frac{\sum_{i\in C}\sum_{j\notin C}\tilde{A}_{ij}^{(W)}}{\sum_{i\in C}\sum_{j\notin V}\tilde{A}_{ij}^{(W)}} \tag{3}$$

$$I_w(C) = 1 - \emptyset_w(C) \tag{4}$$

A larger $I_w(C)$ indicates a higher fraction of within-community flow; in this study we retained only communities with $I_w(C) \geq 0.8$.

Since windows overlap and temporal scales are nested, conceptually similar communities may recur across windows. We therefore enforced membership distinctness using Jaccard similarity $J$ between any two candidate communities $C_a$ and $C_b$:

$$J(C_a, C_b) = \frac{|V(C_a) \cap V(C_b)|}{|V(C_a) \cup V(C_b)|} \tag{5}$$

From coarse to fine scales, we kept the coarser-scale instance and discarded finer-scale near-duplicates that exceeded the threshold of 0.20. Finally, we linked communities across adjacent windows by maximizing Jaccard overlap and designated a community as persistent if it appeared in three or more successive windows. The resulting set of persistent mobility communities is thus (1) grounded in the OD networks defined in Section 3.3.1; (2) optimized by Louvain algorithm at multiple temporal scales; (3) filtered for internal cohesion via inverse conductance; (4) stabilized over time via Jaccard-based tracking. These communities provide the structural basis for the community-level metrics in Section 3.3.4 and the analyses in Section 4.

### 3.3.4 Community-level metrics

- **Hazard-exposure heterophily**

Hazard-exposure heterophily describes the extent to which mobility connections link areas experiencing different levels of hazard impact. In this study, it reflects whether residents in more severely affected tracts maintained or formed mobility ties to less-affected tracts during the managed power outage. Such cross-impact connections can provide access to resources, shelter, and services that may not be available within a community's own high-impact zone.

Here, we first estimated the outage impact level for each census tract in Harris County using the percent change in activity density between the baseline period and the impact period. A value of –75%, for example, indicates that the activity density during the evaluation period is 75% lower than its baseline level. Based on the minimum percent change observed during the outage, we classify tracts into three categories: high-impact (–100%, indicating no Mapbox-detected activity at the lowest point), moderate-impact (between –100% and –75%), and low-impact (> –75%). Hazard-exposure heterophily is expressed in the composition of each detected mobility community: communities that include a mix of high-, moderate-, and low-impact tracts are considered more heterophilous in hazard exposure. These mixed-impact structures suggest that mobility flows spanned areas with different outage severities, potentially facilitating redistribution of people and resources during Winter Storm Uri.

- **Socio-demographic homophily**

Socio-demographic homophily refers to the tendency for individuals to form and maintain stronger connections with others who share similar social and demographic characteristics, such as income level, race, or ethnicity [33, 34]. In the context of hazard response, higher socio-demographic homophily within mobility communities may indicate that population movements during disruption remain concentrated among socially similar groups, which can influence information exchange, mutual aid, and access to recovery resources.

To quantify socio-demographic homophily in the communities identified during Winter Storm Uri, we examine the distribution of median household income, percentage of Black population, and percentage of Hispanic population within each community. We use the interquartile range (IQR), the difference between the 75th and 25th percentiles, as a measure of statistical dispersion. A smaller IQR indicates a narrower spread of values, implying that community members are more socio-demographically similar. Conversely, a larger IQR suggests greater internal diversity in these attributes. By comparing the IQRs of each community to county-wide benchmarks, we assess whether the communities formed during the outage were more or less socio-demographically homogeneous than Harris County as a whole, providing insight into the role of social similarity in shaping mobility patterns under extreme weather conditions.

- **Social-connectedness strength**

Social-connectedness strength captures the extent to which census tracts within a mobility community are socially tied to each other, as opposed to external areas. We measure this using the SCI at the census tract level, which quantifies the relative intensity of Facebook friendship links between two geographic areas. In this study, each geographic unit in the SCI corresponds to a census tract. Higher SCI values indicate a greater density of social ties between the two tracts, suggesting stronger channels for communication, trust, and potential resource sharing.

To evaluate internal connectedness for each identified mobility community $c$, we compute the SCI fraction: the proportion of all social ties involving members of the community that are internal to it:

$$SCIF(c) = \frac{\sum SCI_{i,j}}{\sum SCI_{i,k}}, i, j \in CT_c, k \in CT_{all} \quad (6)$$

where, $SCIF(c)$ is the fraction of the *SCI* needed to understand the social connectedness strength within community $c$; $SCI_{i,j}$ is the *SCI* between census tracts $i$ and $j$ where both $i$ and $j$ are in $CT_c$, census tracts in community $c$, and $SCI_{i,k}$ is the *SCI* between census tracts $i$ and $k$, where $k$ belongs to $CT_{all}$, all census tracts in networks.

A higher $SCIF(c)$ indicates that a greater share of a community's social ties is internal, implying a more tightly knit social structure. Such communities may be better positioned to mobilize collective action and share resources during extreme weather disruptions, whereas communities with lower internal connectedness may be more reliant on external networks for support.

# 4. Results

## 4.1 Detected communities based on human mobility networks

Applying the dynamic community detection procedure described in Section 3.3.3 to the February 2021 human mobility networks yielded 89 distinct communities across all temporal windows (Figure 3a). From these, we identified 30 persistent communities (Figure 3b) that contained more than ten census tracts and persisted for at least three consecutive days, specifically from February 15 to 17, 2021, encompassing the peak of Winter Storm Uri and the associated managed power outages. Figure 3b maps the spatial distribution of these 30 communities across Harris County. A zoomed view of the Houston downtown area is provided in Figure 3c, illustrating finer-scale community boundaries in the urban core. The detected communities generally comprise census tracts that are spatially contiguous, which is consistent with expectations: during extreme weather events with hazardous travel conditions, individuals are less likely to undertake long-distance trips, leading to geographically clustered mobility communities.

Notably, most identified communities are concentrated in the southern and eastern portions of Harris County, while no persistent communities were detected in the northwest. For subsequent analyses, we focus on these 30 persistent communities to examine their hazard-exposure heterophily, socio-demographic homophily, and internal social-connectedness strength.

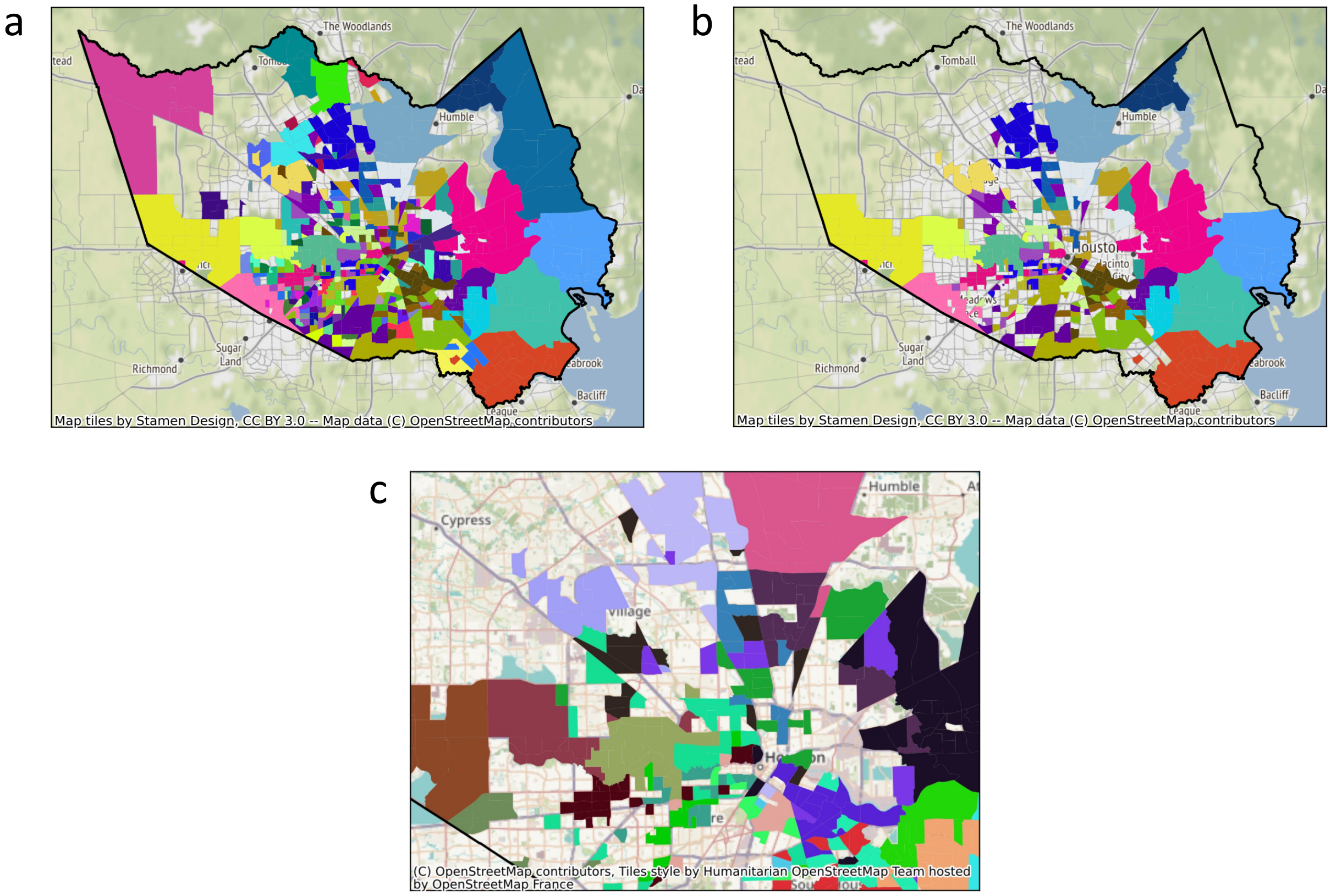

**Figure 3.** (a) Locations of the 89 communities detected across all temporal windows. (b) The 30 persistent communities retained for analysis. (c) Enlarged view of the Houston downtown area showing community boundaries.

### 4.2 Community characteristics during Winter Storm Uri

#### 4.2.1 Hazard-exposure heterophily

Figure 4 demonstrates the percentages of census tracts having high-, moderate-, and low- impact levels of the managed power outages for the identified communities during Winter Storm Uri and their corresponding locations. The average percentage of the high-, moderate-, and low-impact census tracts in the 30 identified communities are 11.9%, 22.9%, and 65.2%, respectively. In particular, the minimum percentage of the low-impact census tracts among the 30 identified communities is 30%, and the minimum percentage of the summation of the medium- and low-impact census tracts is 70%. In other words, the communities detected from the human mobility networks during Winter Storm Uri consist of at least 30% of low-impact areas and 70% of the combination of medium- and low-impact areas. We note that the communities were identified from the human mobility network during the impact period, and the human mobility network represents the connections between people's home census tracts and census tracts where people stay more than two hours with their travel counts. That is said, the identified communities, where people have more in-community movements than the whole network, have at least 30% of the census tracts are low-impact areas that can serve as relocation destinations for people living in high-impact areas to cope with the power outage impacts. Also, the results show that the presence of hazard-exposure heterophily in communities enabled people to temporarily relocate among census tracts. This result highlights the significance of ensuring hazard-exposure heterophily within a community in coping with hazard events so that people living in high-impact areas can seek support from connections in low-impact areas. Thus, hazard-exposure heterophily is a key characteristic in shaping communities in human networks during hazard response.

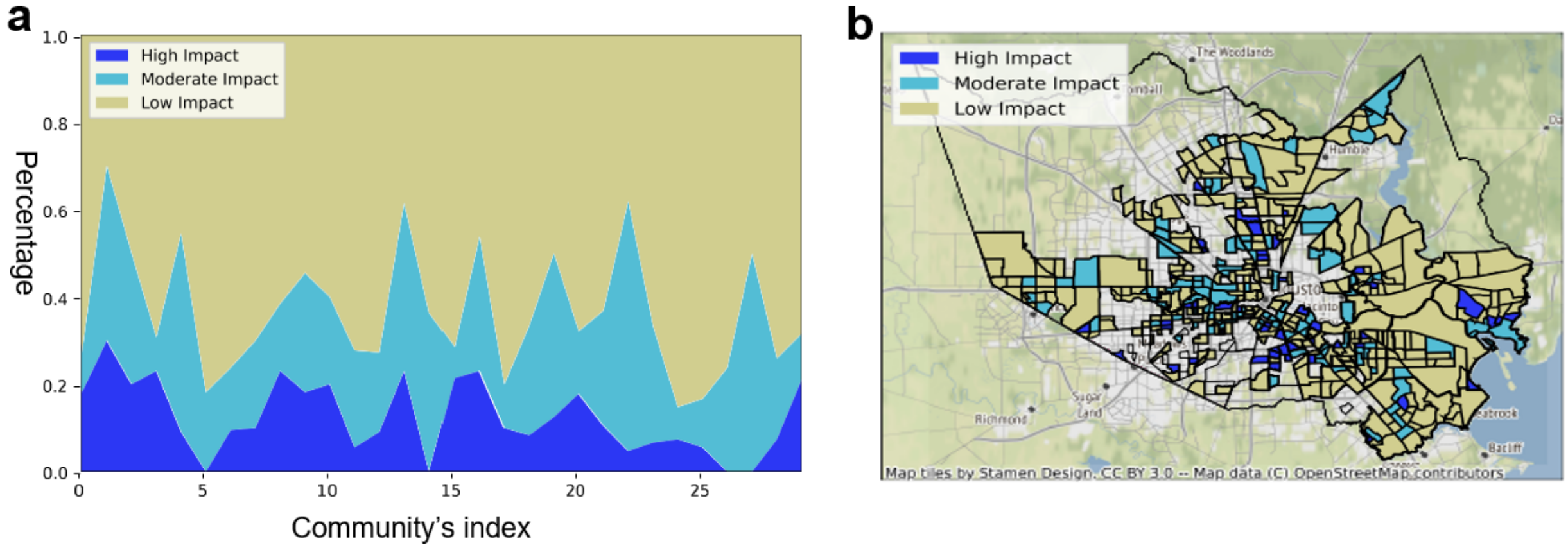

**Figure 4.** (a) The percentages of census tracts having high, moderate, and low levels of the managed power outages impacts for the identified communities during Winter Storm Uri. (b) Spatial distribution of census tracts having high, moderate, and low levels of the managed power outages impacts for the identified communities during Winter Storm Uri. The x axis of subplot (a) is the index of each identified community, and the y axis is the percentages of census tracts having high, moderate, and low levels power outages in the corresponding community.

#### 4.2.2 Socio-demographic homophily

In addition to the impact level and associated heterophily within the identified communities, their socio-demographic homophily was examined. People tend to form communities with people having similar identities, such as income levels [33, 34]. Therefore, it is important to examine extent of socio-demographic homophily in communities formed during hazard response. As shown in Figure 5, we calculated the IQR for all identified communities and Harris County to understand the similarity of the socio-demographic attributes within communities in terms of income, race, and ethnicity. Figure 5a shows the comparison of the IQR between the communities (bars) and Harris County (dashed line) based on their income distributions. During the managed power outage, 22 out of 30 identified communities have lower IQR of income compared to the IQR in Harris County, which is $43K USD. In addition, more than half of the communities have the IQR of income less than $20K USD, which means the income levels within these identified communities are similar.

Besides the IQR of income distributions in communities, Figures 5b and 5c demonstrate the comparison of the IQR between the communities and Harris County based on the percentage of Hispanic and Black populations. In Figure 5b, the IQR of Hispanic population percentage in all communities is less than the IQR of it in Harris County (40.22%). Also, during the winter storm, 21 out of 30 identified communities have lower IQR of Black population percentage compared to the IQR of values in Harris County (22.2%). The results show that most of the identified communities during the managed power outage have relatively similar socio-demographic characteristics compared to Harris County in its entirety in terms of income, race, and ethnicity. This result highlights that socio-demographic homophily plays an important role in the formation of communities during hazard response. Hence, it is crucial for operators to ensure that not all census tracts with similar characteristics are affected at the same time, which may result in people unable to find resources to effectively cope with hazard impacts.

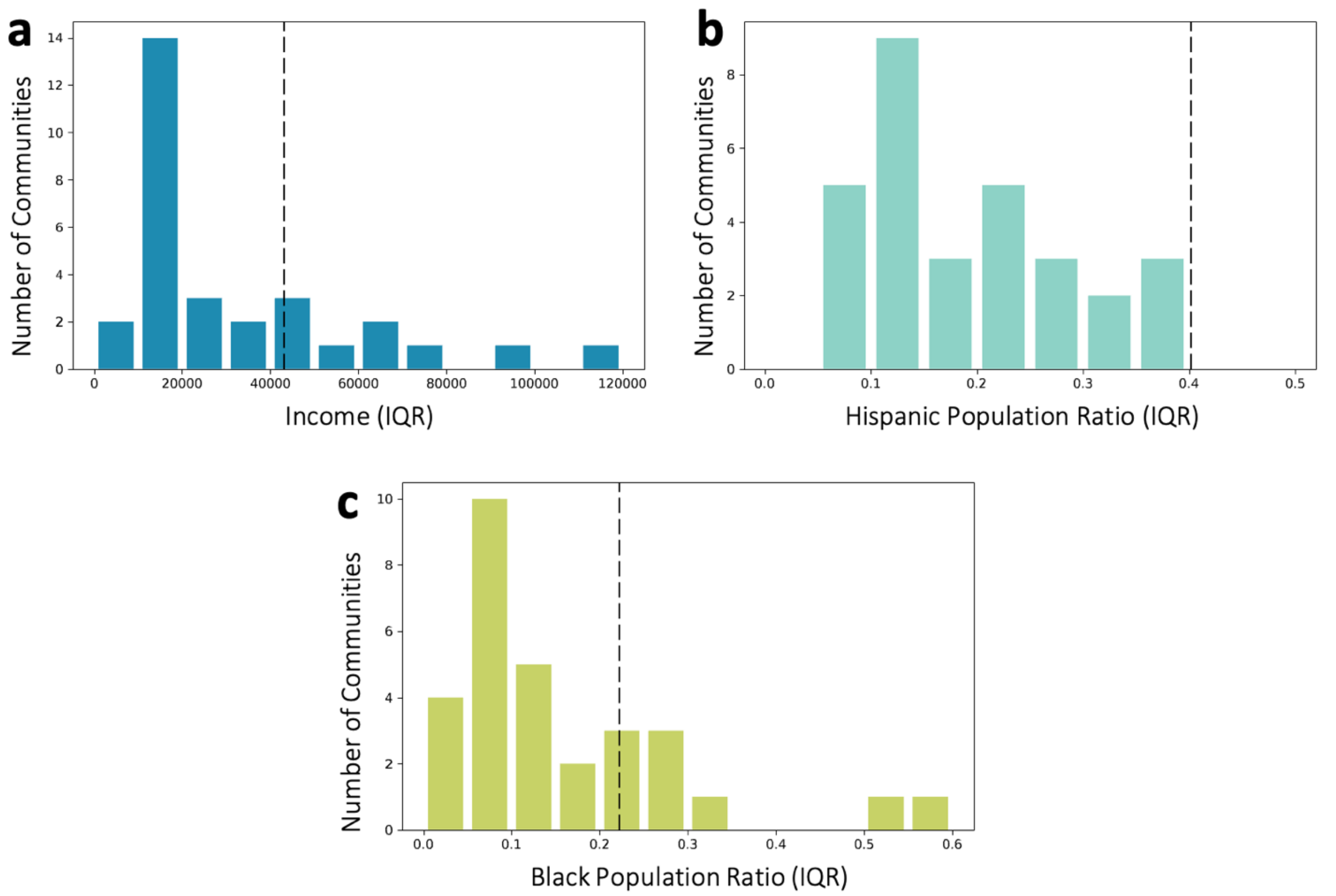

**Figure 5.** (a) The interquartile range of income in the identified communities compared with their interquartile ranges in Harris County. (b) The interquartile range of the ratio of Hispanic population in the identified communities compared with their interquartile ranges in Harris County. (c) The interquartile range of and Black population in the identified communities compared with their interquartile ranges in Harris County. The dashed black lines show the IQR for the entire Harris County.

### 4.2.3 Social-connectedness strength

The third characteristic examined was the strength of social connectedness among spatial areas within communities. Social connectedness among populations of spatial areas is also one of the fundamental characteristics in formation of communities. In this study, we used Meta's Social Connectedness Index to understand the social ties of different census tracts within the identified communities. We calculated a fraction of the SCI to understand the social connectedness strength within the identified communities. The larger the fraction of the SCI of a community, the more inner social connectedness it has. Figure 6 shows the results of the fraction of the SCI for all identified communities. During the managed power outage, 24 out of 30 identified communities had fractions greater than 0.8, which indicates that the social connectedness within the identified communities is more prominent than the social connectedness outside of the communities. This result confirms the importance of social connectedness in the formation of communities within human mobility networks during hazard response. Therefore, if census tracts with high social connectedness experience great impacts (such as extensive power outages) simultaneously, their ability to cope with the impacts diminishes due to the loss of resourceful connections in less impacted areas.

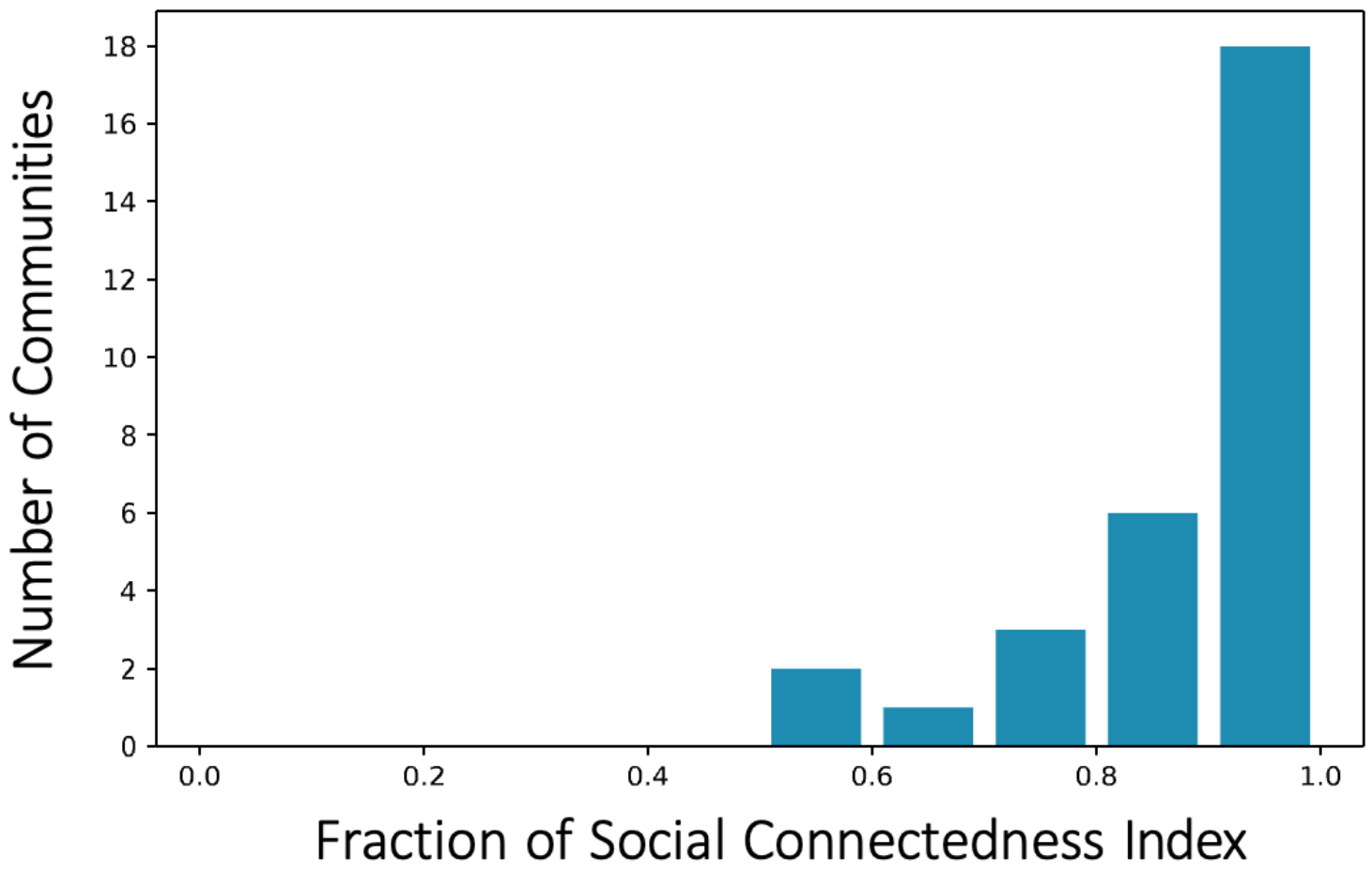

**Figure 6.** Fraction of social connectedness index of the identified communities.

### 4.3 Overflow human mobility movements

In addition to the analysis results related to the three important characteristics, we also found that the proportion of high- and moderate-impact areas within a community is a key factor determining population response. Figure 7 shows the relationship between the fraction of the mobility links from high- and moderate-impact areas to low-impact areas (H/M-to-L links) and the percentage of high- and moderate-impact areas within communities. The fraction of the in-community H/M-to-L links was calculated as the sum of the weighted links between high- and moderate- to low-impact areas within a community divided by the sum of all weighted links between high- and moderate- to low-impact areas where the high- and moderate-impact areas are in the community. A higher fraction of the H/M-to-L links indicates that a community has more in-community links from high- and moderate-impact to low-impact areas, which means that most of the people affected in the community looked for resources from their social connections in the same community. Yet a lower fraction of the H/M-to-L links means that more people living in high- and moderate-impact areas reach out to their social connections outside of the community. The comparison of the fraction of in-community H/M-to-L links to the percentage of high- and moderate-impact areas within communities is shown in Figure 7 with a fitted exponential function. Although the $R^2$ value is 0.31, the fitted curve still shows a relationship between the fraction of in-community H/M-to-L links and the percentage of high- and moderate-impact areas within communities. Specifically, when the percentage of high- and moderate-impact areas in a community is higher, the fraction of the in-community H/M-to-L links tends to be lower. On the other hand, the communities having relatively higher fraction of the in-community H/M-to-L links are likely to have lower percentage of high- and moderate-impact areas in the communities. The result implies that the fraction of high- and moderate-impact areas determines the in-community response capacity. The increase in the fraction triggers out-of-community movements (which may require longer-distance movements or weak social connections). These overflow movements across communities are due mainly to the imbalance in hazard-exposure heterophily of communities.

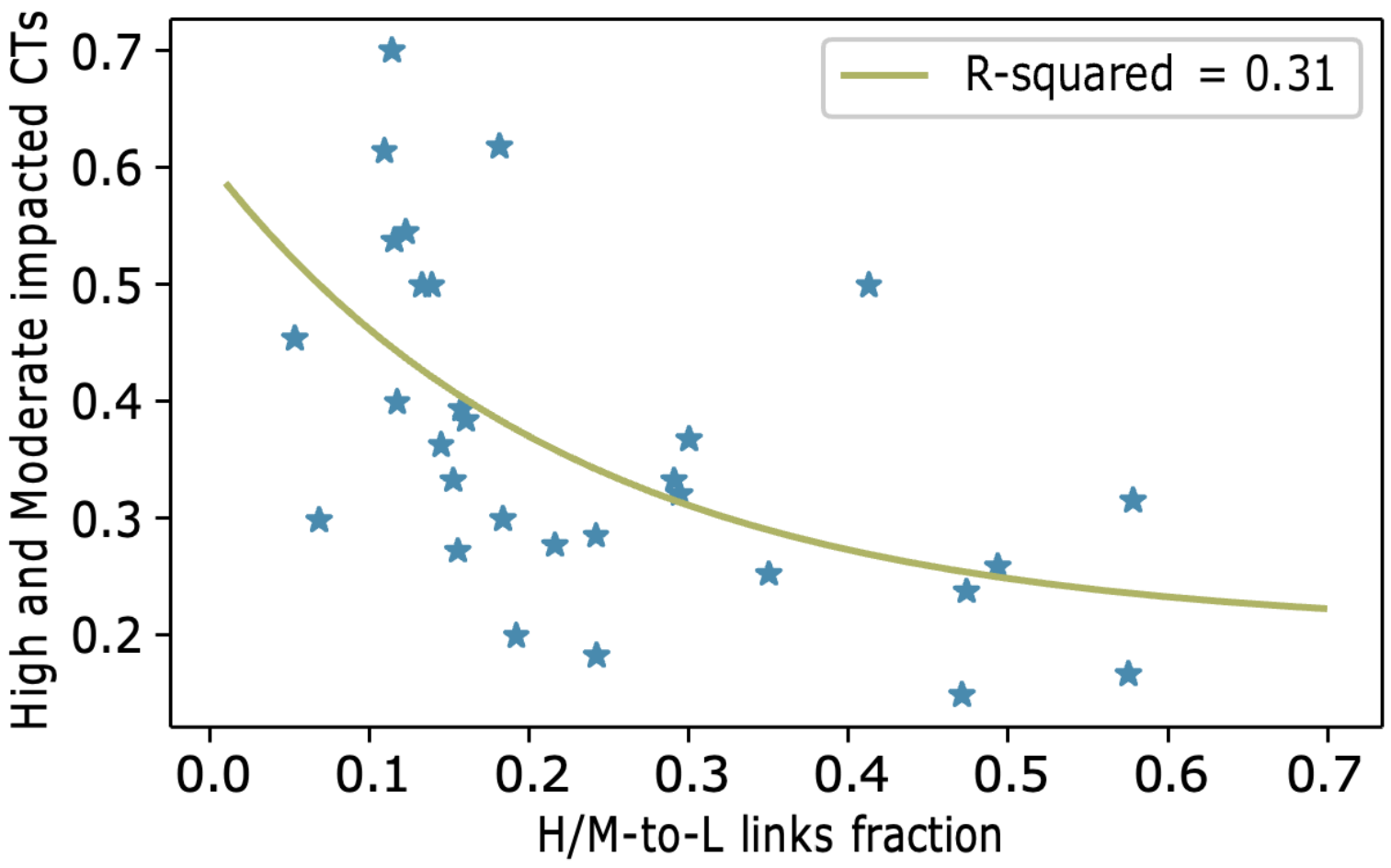

**Figure 7.** Relationship of the fraction of in-community connections from high- and moderate-impact areas to low-impact areas and the percentage of high- and moderate-impact areas within communities. The curve shows an exponential fit.

## 5. Discussion and concluding remarks

Characterizing human network dynamics in response to hazards is essential for better response and faster recovery from impacts. In this study, we examined communities formed in human mobility networks in response to hazards in the context of the 2021 Winter Storm Uri in Harris County, Texas, and its associated managed power outage. Community formation in socio-spatial networks is a mechanism through which populations cope with and respond to hazards and their impacts. We implemented community detection techniques to identify communities based on population responses to the managed power outage event. Then, we examined three important characteristics that shape the formation of communities in response to hazards within the identified communities: heterophilic hazard exposure, homophilic socio-demographics, and strong social connectedness within communities. The analysis results reveal important heterophilic and homophilic characteristics shaping community formation in human mobility networks.

Understanding population responses in mobility networks offers a unique lens for revealing how communities reorganize in real time to cope with hazard impacts. The spatial-temporal structures embedded in these networks capture not only the origin-destination relationships of movement but also the underlying social, demographic, and infrastructural constraints that shape those movements. Our results show that hazard-exposure heterophily (i.e., the coexistence of high-, moderate-, and low-impact tracts within the same mobility community) facilitated potential short-distance relocation during the outage, creating opportunities for affected residents to access resources without leaving their established community boundaries. This finding aligns with prior research on network heterophily as a facilitator of adaptive capacity [44, 51-53] but extends it by providing quantitative evidence at the census-tract scale during a cold-weather, power-outage hazard event that has been far less studied.

In contrast, the strong socio-demographic homophily observed in most communities suggests that relocation and support flows occurred largely within socially similar groups. This pattern echoes earlier findings that racial and income minorities tend to move within socially homogeneous clusters [54-56], but our analysis shows this effect persisting even when heterophilous exposure patterns were present. This suggests that socio-demographic similarity remains a dominant structuring force in mobility networks during crises, potentially limiting the reach of cross-group support unless pre-existing heterogeneous ties exist.

The third characteristic, social-connectedness strength, was high in most communities, indicating that many of the mobility flows during the outage were embedded in pre-existing social networks rather than formed opportunistically in response to the hazard. This reinforces the idea that social infrastructure (e.g., networks of trust, reciprocity, and communication) plays a critical role in shaping physical movement during disasters [57, 58]. It also implies that interventions aimed at fostering inter-community connections before hazard events could expand relocation and resource-sharing options during crises.

In addition, we identified overflow movements phenomenon, in which residents in heavily impacted communities increasingly sought resources outside their own communities when the internal proportion of high- and moderate-impact tracts exceeded a certain level. This behavior points to the critical role of hazard-exposure balance in sustaining in-community response capacity and preventing displacement into less familiar social and geographic environments. By jointly analyzing these structural characteristics, our study contributes new evidence on the interplay between network composition, social structure, and adaptive mobility during extreme weather events, providing a more nuanced understanding of how human mobility networks operate as both a reflection of social structures and a mechanism of hazard response.

Our findings offer valuable implications for emergency management planning and response. Socio-demographic similarities and pre-existing social connections are central to how communities mobilize during hazards, suggesting that response strategies should work with these inherent dynamics. Communities with strong internal social ties can be prioritized as hubs for localized resource distribution, rapid information dissemination, and mutual aid coordination. Conversely, results on hazard-exposure heterophily and overflow movements indicate that communities with a high share of severely affected tracts may need targeted external support to avoid reliance on distant or weaker networks. Integrating mobility network analysis into preparedness planning can help identify such vulnerabilities in real time, enabling more effective, equitable, and context-sensitive interventions that strengthen both in-community coping capacity and cross-community support pathways.

There are also some limitations on this work. First, the availability and reliability of human mobility data, which depend on infrastructures such as electricity and communication networks, might be compromised during such hazardous events. A future direction for this research is to conduct a detailed comparison of the data size and availability before, during, and after the event to justify the usability and reliability of the mobility dataset. This comparison will provide insights into the extent of data gaps caused by power outages and enhance the robustness of our findings. Second, due to the availability of the data, we are not able to accurately calculate the number of device count fluctuations caused by evacuation or migration. Future research could benefit from integrating additional data sources or methods to better distinguish between different types of travel motivations. Third, while it was reported that more than 90% of residents in Harris County lost power during Winter Storm Uri [14, 41], the specific causes of these outages, whether due to distribution system failures or broader transmission system issues, require further investigation. This distinction is important for understanding the full context of the mobility network findings. Future studies should aim to delineate the exact causes of power outages to better relate mobility impacts to specific infrastructure failures.

## Data Availability

The data that support the findings of this study are available from Spectus and Meta Social Connectedness Data, but restrictions apply to the availability of these data. The data can be accessed upon request submitted to Spectus and Meta Data for Good Program. Other data used in this study are all publicly available.

## Code Availability

The code that supports the findings of this study is available from the corresponding author upon request.

## Acknowledgements

The authors would like to acknowledge the funding support from the National Science Foundation CAREER Award under grant number 1846069. The authors would also like to acknowledge Meta's Data for Good Program for providing the social connectedness data. Any opinions, findings, conclusions, or recommendations expressed in this research are those of the authors and do not necessarily reflect the view of the funding agencies.

## Author Contributions Statement

C.-C.L. designed the analysis. C.-C.L., J.-W.M, K.Y., S.N., and X.X. analyzed the results. C.-C.L., J.-W.M, K.Y., prepared the main manuscript text. A.M supervised the analysis and revised the main manuscript text. All authors reviewed the manuscript.

## Competing Interests

The authors declare no competing interests.